\documentclass{article}

\PassOptionsToPackage{numbers, sort&compress}{natbib}

\usepackage[preprint]{neurips_2021}

\usepackage[utf8]{inputenc} 
\usepackage[T1]{fontenc}    
\usepackage{hyperref}       
\usepackage{url}            
\usepackage{booktabs}       
\usepackage{amsfonts}       
\usepackage{nicefrac}       
\usepackage{microtype}      
\usepackage{xcolor}         

\usepackage{amssymb}
\usepackage{amsthm}
\usepackage{amsmath}
\usepackage{graphicx}
\usepackage{multirow}
\usepackage{enumerate}

\usepackage{appendix}

\usepackage{subcaption}

\renewcommand{\vec}[1]{\boldsymbol{\mathbf{#1}}}

\DeclareMathOperator{\EX}{\mathbb{E}}
\newcommand{\model}{NVC-Net}

\title{NVC-Net: End-to-End Adversarial Voice Conversion}

\author{%
  Bac Nguyen \\
  Sony Europe B.V.\\
  \texttt{Bac.NguyenCong@sony.com} \\
  \And
  Fabien Cardinaux \\
  Sony Europe B.V. \\
  \texttt{Fabien.Cardinaux@sony.com}
}

\begin{document}

\maketitle

\begin{abstract}
Voice conversion has gained increasing popularity in many applications of speech synthesis. The idea is to change the voice identity from one speaker into another while keeping the linguistic content unchanged. Many voice conversion approaches rely on the use of a vocoder to reconstruct the speech from acoustic features, and as a consequence, the speech quality heavily depends on such a vocoder. In this paper, we propose \model{}, an end-to-end adversarial network, which performs voice conversion directly on the raw audio waveform of arbitrary length. By disentangling the speaker identity from the speech content, \model{} is able to perform non-parallel traditional many-to-many voice conversion as well as zero-shot voice conversion from a short utterance of an unseen target speaker. Importantly, \mbox{\model{}} is non-autoregressive and fully convolutional, achieving fast inference. Our model is capable of producing samples at a rate of more than 3600 kHz on an NVIDIA V100 GPU, being orders of magnitude faster than state-of-the-art methods under the same hardware configurations. Objective and subjective evaluations on non-parallel many-to-many voice conversion tasks show that \model{} obtains competitive results with significantly fewer parameters.
\end{abstract}

\section{Introduction}

Voice conversion consists in changing the speech of a source speaker in such a way that it sounds like that of a target speaker while keeping the linguistic information unchanged~\cite{8553236,9262021}. Typical applications of voice conversion include speaker conversions, dubbing in movies and videos, speaking aid systems~\cite{NAKAMURA2012134}, and pronunciation conversion~\citep{Kaneko2017}. Despite much success in this field, several issues remain to be addressed for voice conversion. 

Early voice conversion methods require parallel training data, which contain utterances of the same linguistic content spoken by different speakers. A drawback of these methods is their sensitivity to misalignment between the source and target utterances~\citep{5485203}. Time alignment and manual correction are typically performed as a pre-processing step to make these methods work more reliably~\cite{helander2008impact}. Nevertheless, collecting parallel training data is still a time-consuming and can be even an infeasible task in many situations~\citep{Lorenzo-Trueba2018}. This has motivated growing research interest in developing voice conversion methods on non-parallel training data by using generative adversarial networks (GANs)~\citep{NIPS2014_5ca3e9b1} and its variants. Although GAN-based methods provide a promising solution by learning a mapping through non-parallel training data~\cite{8639535,8553236,kaneko2017parallel,Kaneko2019}, designing such a method that yields a competitive performance compared to parallel methods still remains to be solved.

Another common problem in voice conversion is modeling the raw audio waveform. A good model should capture the short and long-term temporal dependencies in audio signals, which is not trivial due to the high temporal resolution of speech (\textit{e.g.}, 16k samples per second). A simple solution is to reduce the temporal resolution into a lower-dimensional representation (\textit{e.g.}, acoustic features). In such a case, one needs a vocoder that reconstructs the waveform representation from acoustic features. Traditional vocoders like WORLD~\citep{Masanori} or STRAIGHT~\cite{10.1016/S0167} often introduce artifacts (metallic sounds). To overcome this issue, more sophisticated neural vocoders such as autoregressive~\cite{vandenOord_2016,vandenOord_2016} and non-autoregressive~\cite{8683143,NEURIPS2019_6804c9bc,9053795} models have been proposed. Autoregressive models such as WaveNet~\cite{vandenOord_2016} and WaveRNN~\cite{pmlr-v80-kalchbrenner18a} are slow at inference, therefore, they are not well suited for real-world applications. Although non-autoregressive models like WaveGlow~\cite{8683143} yield faster inference, the waveform reconstruction  still dominates the computational effort as well as the memory requirements. Lightweight vocoders like MelGAN~\cite{NEURIPS2019_6804c9bc} or Parallel WaveGAN~\cite{9053795} can alleviate the memory issue, however, the feature mismatch problem between the training and inference remains unsolved. Indeed, as observed by~\citet{Wu2018}, the feature mismatch problem may cause the vocoder to produce noisy or collapsed speech in the output, especially when training data are limited.

While many research efforts have focused on converting audio among speakers seen during training, little attention has been paid to the problem of converting audio from and to speakers unseen during training (\textit{i.e.}, zero-shot voice conversion).  \citet{pmlr-v97-qian19c} proposed to use a pre-trained speaker encoder on a large data set with many speakers. With carefully designed bottleneck layers, the content information can be disentangled from the speaker information, allowing it to perform zero-shot voice conversion. However, under the context of limited data and computational resources, the speaker encoder cannot generalize well on unseen speakers since the speaker embeddings tend to be scattered.

In this paper, we tackle the problem of voice conversion from raw audio waveform using adversarial training in an end-to-end manner. The main contributions of this paper are summarized as follows.
\begin{enumerate}[(i)]
    \item We introduce \model{}, a many-to-many voice conversion method trained adversarially, which does not require parallel data. Our method aims to disentangle the speaker identity from the speech content. To the best of our knowledge, this is the first GAN-based method that explicitly performs disentanglement for voice conversion directly on the raw audio waveform.
    \item Unlike other GAN-based voice conversion methods, \model{} can directly generate raw audio without training an additional vocoder. Thus, the feature mismatch problem when using an independent vocoder can be avoided. The absence of a vocoder also allows \model{} to be very efficient at inference time. It is capable of generating audio at a rate of more than 3600 kHz on an NVIDIA V100 GPU.
    \item \model{} addresses the zero-shot voice conversion problem by constraining the speaker representation. We apply the Kullback-Leibler regularization on the speaker latent space, making the speaker representation robust to small variations of input. This also allows stochastic sampling at generation. During inference, one can randomly sample a speaker embedding from this space to generate diverse outputs or extract the speaker embedding from a given reference utterance.
\end{enumerate}

\section{Related works} \label{sec:relatedwork}
Current voice conversion methods resort to non-parallel training data due to the difficulties in collecting parallel training data. Many of them have been inspired by the recent advances in computer vision. For instance, \citet{7820786} introduced a variational auto-encoder (VAE)-based voice conversion framework by conditioning the decoder on the speaker identity. As an extension, \citet{8718381} used an auxiliary classifier for VAEs forcing the converted audio to be correctly classified. Despite the success, it is well known that conditional VAEs have the limitation of generating over-smoothed audio. This can be problematic as the model may produce poor-quality audio~\cite{8639535}.

An appealing solution to solve this problem is based on GANs~\citep{NIPS2014_5ca3e9b1}  since they do not make any explicit assumption about the data distribution. They learn to capture the distribution of real audio, making the generated audio sounding more natural~\cite{Kaneko2019,NEURIPS2019_6804c9bc,9053795}. First studies were based on CycleGAN~\citep{8553236,8462342,kaneko2017parallel}. The idea is to learn a mapping on acoustic features between the source and target speakers by combining the adversarial loss, cycle-consistency loss, and identity mapping loss. More recently, inspired by the work of~\citet{Choi_2018_CVPR}, StarGAN-VC~\citep{8639535} and its improved version~\citep{Kaneko2019} showed very promising results for voice conversion. \citet{hsu2017voice} demonstrated that GANs could improve VAE-based models. However, despite the success in generating realistic images, building GANs to generate directly high-fidelity audio waveforms is a challenging task~\citep{NEURIPS2019_6804c9bc,donahue2018adversarial,binkowski2019high}. One reason is that audio are highly periodic and human ears are sensitive to discontinuities. Moreover, unlike images, audio often induce high temporal resolution requirements.

In the context of voice conversion, we consider speech as a composition of speaker identity and linguistic content. Disentangling the speaker identity from the linguistic content allows to change the speaker identity independently. There have been several methods leveraging the latent representation of automatic speech recognition (ASR) models for voice conversion~\cite{Liu2018,Zhang2020}. An ASR model is used to extract the linguistic features from the source speech, then convert these features into a target speaker using another speaker-dependent model. These models can perform non-parallel voice conversion, but their performance highly depends on the accuracy of the ASR model. Other disentanglement techniques for voice conversion include autoencoder~\cite{pmlr-v97-qian19c}, vector quantization~\cite{9053854}, and instance normalization~\cite{Chou2019}.

Many of the aforementioned approaches rely on the intermediate representation of speech, such as spectrograms and mel-frequency cepstral coefficients. Although they are capable of producing good perceptual-quality speech, some still need additional supervision to train a robust vocoder. There are only a few methods dealing with the raw audio waveform. For instance, \citet{pmlr-v70-engel17a} proposed a WaveNet-style autoencoder for converting audio between different instruments. More recently, \citet{NEURIPS2019_9426c311} introduced Blow, a normalizing flow network for voice conversion on raw audio signals. Flow-based models have the advantages of efficient sampling and exact likelihood computation, however, they are less expressive compared to autoregressive models.

\section{NVC-Net} \label{sec:main}
\begin{figure}
    \centering
    \includegraphics[width=\textwidth]{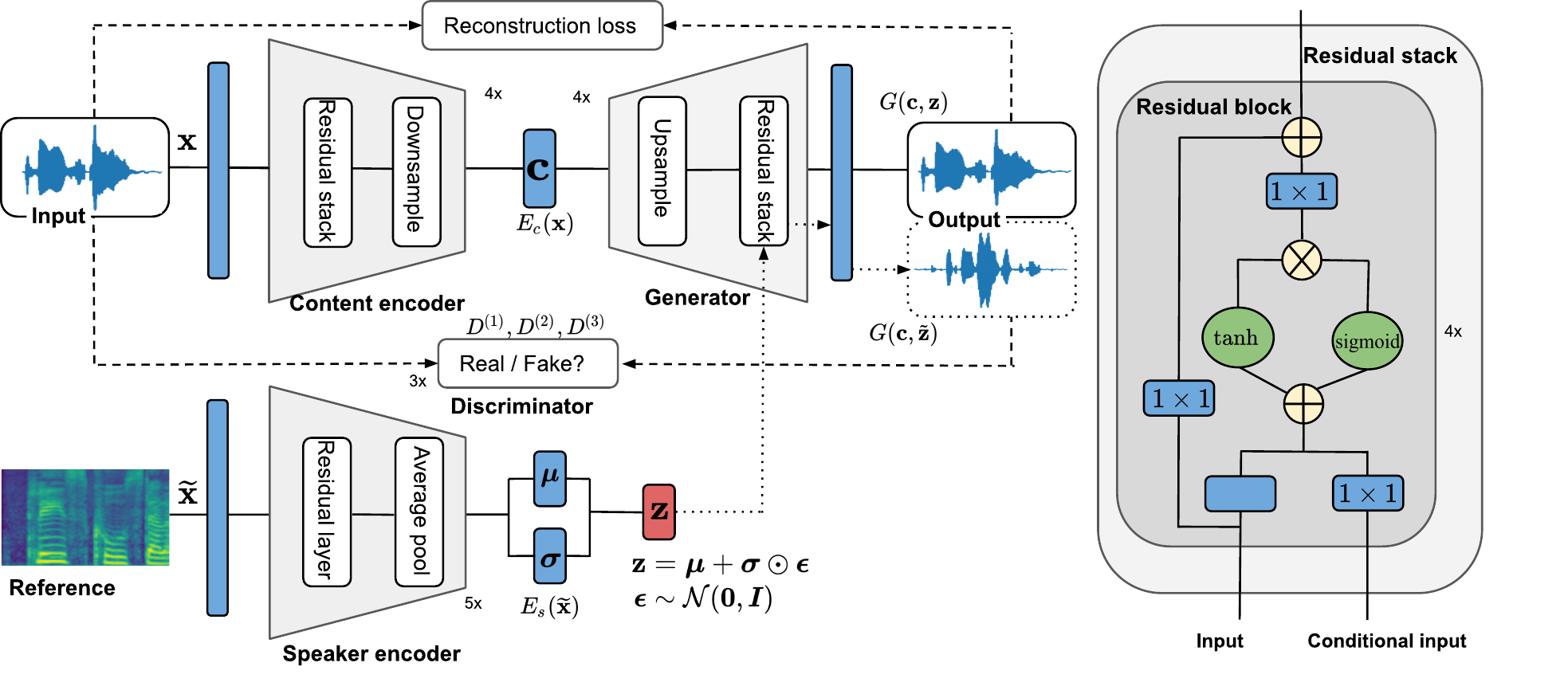}
    \caption{Overview training procedure and network architecture of \model{}. The model consists of a speaker encoder, a content encoder, a generator, and three discriminators. The rectangle filled with blue color denotes a convolutional layer. \model{} produces high-fidelity audio (enforced by the discriminators) matching the input (enforced by the reconstruction loss).}
    \label{fig:architecture}
    \vspace{-2mm}
\end{figure}

Our proposal is designed with the following objectives: (i) reconstructing highly-perceptually-similar audio waveform from latent embeddings, (ii) preserving the speaker-invariant information during the conversion, and (iii) generating high-fidelity audio for a target speaker. In particular, \model{} consists of a content encoder $E_c$, a speaker encoder $E_s$, a generator $G$, and three discriminators $D^{(k)}$ for $k=1,2,3$ which employ on different temporal resolutions of the input. Figure~\ref{fig:architecture} illustrates the overall architecture. We assume that an utterance $\vec{x}$ is generated from two latent embeddings, speaker identity $\vec{z} \in \mathbb{R}^{d_{\text{spk}}}$ and speech content $\vec{c}\in \mathbb{R}^{d_{\text{con}}\times L_{\text{con}}}$, \textit{i.e.}, $\vec{x} = G(\vec{c}, \vec{z})$. The content describes information that is invariant across different speakers, \textit{e.g.}, phonetic and other prosodic information. To convert an utterance $\vec{x}$ from speaker $y$ to speaker $\widetilde{y}$ with an utterance $\widetilde{\vec{x}}$, we map $\vec{x}$ into a content embedding through the content encoder $\vec{c} = E_c(\vec{x})$. In addition, the target speaker embedding $\widetilde{\vec{z}}$ is sampled from the output distribution of the speaker encoder $E_s(\widetilde{\vec{x}})$. Finally, we generate raw audio from the content embedding $\vec{c}$ conditioned on the target speaker embedding $\widetilde{\vec{z}}$, \textit{i.e.}, $\widetilde{\vec{x}} = G(\vec{c}, \widetilde{\vec{z}})$.

\subsection{Training objectives}
Our goal is to train the content encoder $E_c$, the speaker encoder $E_s$, and the generator $G$ that learns a mapping among multiple speakers. In the following, we will explain in detail the training objectives.

\textbf{Adversarial loss.} To make synthesized voices indistinguishable from real voices, the adversarial losses for the discriminator $\mathcal{L}_{\text{adv}}(D^{(k)})$ and for the generation $\mathcal{L}_{\text{adv}}(E_c,E_s,G)$ are defined as

\begin{equation}
\begin{aligned}
    \mathcal{L}_{\text{adv}}(D^{(k)}) &= -\EX_{\vec{x}, y}\Big[ \log D^{(k)}(\vec{x})[y]\Big]  - \EX_{\vec{c}, \widetilde{\vec{z}}, \widetilde{y}}\left[ \log\big(1 - D^{(k)} (G(\vec{c}, \widetilde{\vec{z}}))[ \widetilde{y}]\big)\right],\\
     \mathcal{L}_{\text{adv}}(E_c,E_s,G) &=   \sum_{k=1}^3 \EX_{\vec{c}, \widetilde{\vec{z}}, \widetilde{y}}\left[\log\big(1- D^{(k)}(G(\vec{c}, \widetilde{\vec{z}}))[\widetilde{y}] \big)\,\right] \,.
\end{aligned}
\label{eq:adv}
\end{equation}

The encoders and generator are trained to fool the discriminators, while the discriminators are trained to solve multiple binary classification tasks simultaneously. Each discriminator has multiple output branches, where each branch corresponds to one task. Each task consists in determining, for one specific speaker, whether an input utterance is real or converted by the generator.  The number of branches is equal to the number of speakers. When updating the discriminator $D^{(k)}$ for an utterance $\vec{x}$ of class $y$, we only penalize if its $y$-th branch output $D^{(k)}(\vec{x})[y]$ is not correct, while leaving other branch outputs untouched. In our implementation, a reference utterance $\widetilde{\vec{x}}$ for a source utterance $\vec{x}$ is taken randomly from the same mini-batch.

\textbf{Reconstruction loss.} When generating the audio, we force the generator $G$ to use the speaker embedding by reconstructing the input from the content and speaker embeddings. One can use the point-wise loss on the raw waveform to measure the difference between the original and generated audio. However, the point-wise loss cannot correctly capture the differences among them since two perceptually-identical audio might not have the same audio waveform~\cite{engel2020ddsp}. Instead, we use the following feature matching loss~\citep{pmlr-v48-larsen16,NEURIPS2019_6804c9bc},
\begin{align*}
    &\mathcal{L}^{(k)}_{\text{fm}}(E_c, E_s, G) = \EX_{\vec{c},\vec{z},\vec{x}} \left[ \sum_{i=1}^L \frac{1}{N_i}\Big\|D^{(k)}_i(\vec{x}) - D^{(k)}_i\big(G(\vec{c},\vec{z})\big)\Big\|_1 \right]\,,  
\end{align*}
where $D^{(k)}_i$ denotes the feature map output of $N_i$ units from the discriminator $D^{(k)}$ at the $i$-th layer, $\|.\|_1$ denotes the $\ell_1$-norm, and $L$ denotes the number of layers. The difference is measured through feature maps of the discriminators~\citep{pmlr-v48-larsen16,NEURIPS2019_6804c9bc}. To further improve the fidelity of speech, we add the following spectral loss
\begin{align*}
 \mathcal{L}^{(w)}_{\text{spe}}(E_c, E_s,G)& = \EX_{\vec{c},\vec{z},\vec{x}} \left[\big\|\theta(\vec{x}, w) - \theta\big(G(\vec{c},\vec{z}), w\big)\big\|^2_2 \right]\,,  
\end{align*}
where $\theta(., w)$ computes the log-magnitude of mel-spectrogram with a FFT size of $w$ and $\|.\|_2$ denotes the $\ell_2$-norm. Measuring the difference between the generated and ground-truth signals on the spectral domain has the advantage of increased phase invariance.
Following~\citet{engel2020ddsp}, the spectral loss is computed at different spatial-temporal resolutions using different FFT sizes $w\in \mathcal{W}= \{2048, 1024, 512\}$. Finally, the total reconstruction loss is the sum of all spectral losses and the feature matching losses, \textit{i.e.},
\begin{align}
    \mathcal{L}_{\text{rec}}(E_c, E_s, G) = \sum_{k=1}^3\mathcal{L}^{(k)}_{\text{fm}}(E_c, E_s, G) + \beta\sum_{w\in \mathcal{W}} \mathcal{L}^{(w)}_{\text{spe}}(E_c,E_s,G)\,, \label{eq:recon}
\end{align}
where $\beta\ge0$ is a weighing term. In our experiments, we simply set $\beta=1$.

\textbf{Content preservation loss.} To encourage that the converted utterance preserves the speaker-invariant characteristics of its input audio, we minimize the following content preservation loss
\begin{align}
\mathcal{L}_{\text{con}}(E_c, G) = \EX_{\vec{x}, \widetilde{\vec{z}}}\left[ \big\|E_c(\vec{x}) - E_c\big(G(E_c(\vec{x}), \widetilde{\vec{z}})\big) \big\|^2_2\right]. \label{eq:content}
\end{align}
We encourage the content encoder $E_c$ to preserve the essential characteristics of an input $\vec{x}$ while changing its speaker identity~$\widetilde{\vec{z}}$ during the conversion. There are two potential benefits of adding this loss. First, it allows cycle conversion in the sense that if we convert, \textit{e.g.}, an utterance from a speaker A to a speaker B and then convert it back from B to A, we should obtain the original utterance provided that the reconstruction is also minimized. Second, minimizing Eq.~(\ref{eq:content}) also results in disentangling the speaker identity from the speech content. It can be seen that if the content embeddings of utterances from different speakers are the same, the speaker information cannot be embedded in the content embedding. Unlike previous works~\cite{Choi_2018_CVPR,8639535,Kaneko2019}, we do not perform any domain classification loss on the content embedding, making the training procedure simpler. Interestingly, we observe that the numerical value of $\ell_2$-norm in Eq.~(\ref{eq:content}) can be influenced by the scale of the output from $E_c$. By simply scaling down any $E_c(\vec{x})$, the content preservation loss will be reduced. As a result, the magnitude of content embedding will be relatively small, making the distance between two content embeddings meaningless. To avoid such situation, we regularize the content embedding to have a unit $\ell_2$-norm on the spatial dimension, \textit{i.e.}, $ c_{ij} \leftarrow  c_{ij}/(\sum_k c_{kj}^2)^{1/2}$. In Appendix~\ref{appexdix:ablation}, we conduct an  empirical analysis for this normalization.

\textbf{Kullback-Leibler loss.} To perform stochastic sampling from the speaker latent space, we penalize the deviation of the speaker output distribution from a prior zero-mean unit-variance Gaussian, \textit{i.e.},
\begin{align}
\mathcal{L}_{\text{kl}} (E_s) &= \EX_{\vec{x}} \Big[\mathbb{D}_{\text{KL}}\big(p(\vec{z}|\vec{x}) \| \mathcal{N}(\vec{z}|\vec{0}, \vec{I})\big)\Big]\,, \label{eq:kl}
\end{align}
where $\mathbb{D}_{\text{KL}}$ denotes the Kullback-Leibler (KL) divergence and $p(\vec{z}|\vec{x})$ denotes the output distribution of $E_s(\vec{x})$. Constraining the speaker latent space provides two simple ways to sample a speaker embedding at inference: (i) sample from the prior distribution $\mathcal{N}(\vec{z}|\vec{0}, \vec{I})$ or (ii) sample from  $p(\vec{z} | \vec{x})$ for a reference $\vec{x}$. On one hand, this term enforces the speaker embeddings to be smooth and less scattered, making generalization to unseen samples. On the other hand, we implicitly maximize the lower bound approximation of the log-likelihood $\log p(\vec{x})$ (see our derivation in Appendix~\ref{appendix:loss}).

\textbf{Final loss.} From Eqs.~(\ref{eq:adv}) to (\ref{eq:kl}), the final loss function for the encoders, generator, and discriminators can be summarized as follows
\begin{align*}
    \mathcal{L}(E_c, E_s, G) =& \, \mathcal{L}_{\text{adv}}(E_c, E_s, G) +  \lambda_{\text{rec}}\mathcal{L}_{\text{rec}}(E_c, E_s,G) + \lambda_{\text{con}}\mathcal{L}_{\text{con}}(E_c, G) + \lambda_{\text{kl}}\mathcal{L}_{\text{kl}} (E_s)\,,\\
    \mathcal{L}(D) =&  \sum_{k=1}^3 \mathcal{L}_{\text{adv}}(D^{(k)})\,,
\end{align*}
where $\lambda_{\text{rec}}\ge 0$, $\lambda_{\text{con}}\ge 0$, and $\lambda_{\text{kl}}\ge 0$ control the weights of the objective terms. For our experiments, these hyper-parameters are set to $\lambda_{\text{rec}}=10$, $\lambda_{\text{con}}=10$, and $\lambda_{\text{kl}}=0.02$.

\subsection{Model architectures}
We describe the main components of \model{}. Some structures of \model{} are inherited from MelGAN~\cite{NEURIPS2019_6804c9bc} and WaveNet~\cite{vandenOord_2016}. More details on network configurations are given in Appendix~\ref{appendix:network}.

\textbf{Content encoder.} The content encoder is a fully-convolutional neural network, which can be applied to any input sequence length. It maps the raw audio waveform to an encoded content representation. The content embedding is at 256x lower temporal resolution than its input. The network consists of 4 downsampling blocks, followed by two convolutional layers with the GELU~\cite{hendrycks2016gaussian} activations. A downsampling block consists of a stack of 4 residual blocks, followed by a strided convolution. A residual block contains dilated convolutions with gated-tanh nonlinearities and residual connections. This is similar to WaveNet~\citep{vandenOord_2016}, but here we simply use reflection padding and do not apply causal convolutions. For the content encoder, no conditional input is feed to the residual block. By increasing the dilation in each residual block, we aim to capture long-range temporal dependencies of audio signals as the receptive field increases exponentially with the number of blocks.

\textbf{Speaker encoder.} The speaker encoder produces an encoded speaker representation from an utterance. We assume that $p(\vec{z}|\vec{x})$ is a conditionally independent Gaussian distribution. The network outputs a mean vector $\vec{\mu}$ and a diagonal covariance $\vec{\sigma}^2\vec{I}$, where $\vec{I}$ is an identity matrix. Thus, a speaker embedding is given by sampling from the output distribution, \textit{i.e.}, $\vec{z} \sim \mathcal{N}(\vec{\mu}, \vec{\sigma}^2\vec{I})$. Although the sampling operation is non-differentiable, it can be reparameterized as a  differentiable operation using the reparameterization trick~\cite{kingma:vae}, \textit{i.e.}, $\vec{z} = \vec{\mu} + \vec{\sigma} \odot \epsilon$, where $\epsilon\sim \mathcal{N}(\vec{0}, \vec{I})$. We extract mel-spectrograms from the audio signals and use them as inputs to the speaker encoder. The network consists of 5 residual blocks~\cite{Kaneko2019} with dilated 1D convolutions. We use the Leaky ReLU nonlinarity with a negative slope of 0.2 for activation. Finally, an average pooling is used to remove temporal dimensions, followed by two dense layers, which output the mean and covariance.

\textbf{Generator.} The generator maps the content and speaker embeddings back to raw audio. Our network architecture inherits from MelGAN~\citep{NEURIPS2019_6804c9bc}, but instead of upsampling pre-computed mel-spectrograms, its input comes from the content encoder. More specifically, the network consists of 4 upsampling blocks. Each upsampling is performed by a transposed convolutional layer, followed by a stack of 4 residual blocks with the GELU~\cite{hendrycks2016gaussian} activations. To avoid artifacts, the kernel size is chosen to be a multiple of the stride~\citep{NEURIPS2019_6804c9bc}. Different from MelGAN, our model uses gated-tanh nonlinearities~\citep{vandenOord_2016} in the residual blocks. To perform the conditioning on the speaker identity, we feed a speaker embedding to the residual block as in~\cite{vandenOord_2016} (see also Fig.~\ref{fig:architecture}). The speaker embedding first goes through a $1\times1$ convolutional layer to reduce the number of dimensions to match the number of feature maps used in the dilation layers. Then, its output is broadcast over the time dimension. Finally, we generate the output audio waveform by applying a convolutional layer with the $\tanh$ activation at the end.

\textbf{Discriminator.} The discriminator architecture is similar to that used in MelGAN~\citep{NEURIPS2019_6804c9bc}. In particular, three discriminators with an identical network architecture are applied on different audio resolutions, \textit{i.e.}, downsampled versions of the input with different scales of 1, 2, and 4, respectively. We use strided average pooling with a kernel size of 4 to downsample the audio scales. Differently from MelGAN, our discriminator is a multi-task discriminator~\citep{05cf1}, which contains multiple linear output branches. The number of output branches corresponds to the number of speakers. Each branch is a binary classification task determining whether an input is a real sample of the source speaker or a conversion output coming from the generator. Instead of classifying the entire audio sequences, we also leverage the idea of PatchGANs~\citep{8100115}, which use a window-based discriminator network to classify whether local audio chunks are real or fake.

\subsection{Data augmentation}
It is well known that training GANs with limited data might lead to over-fitting. In order to learn the semantic information instead of memorizing the input signal, we apply a few data augmentation strategies. Human auditory perception is not affected when shifting the phase of the signal by 180 degrees, therefore, we can flip the sign of the input by multiplying with -1 to obtain a different input. Similarly to~\citet{NEURIPS2019_9426c311}, a random amplitude scaling is also performed. In particular, the amplitude of input is scaled by a random factor, drawn uniformly in a range of $[0.25, 1]$. To reduce artifacts, we apply a small amount of temporal jitter (in a range of $[-30, 30]$) to the ground-truth waveform when computing the reconstruction loss in Eq.~(\ref{eq:recon}). Note that these data augmentation strategies do not change the speaker identity or the content information of the signal. Moreover, we introduce a random shuffle strategy, which is applied to the speaker encoder network. First, the input signal is divided into segments of uniformly random lengths from 0.35 to 0.45 seconds. We then shuffle the order of these segments and then concatenate them to form a new input with the linguistic information shuffled in a random order. We observe that the random shuffle strategy helps to avoid that the content information leaks into the speaker embedding, thus achieving better disentanglement.

\section{Experiments} \label{sec:experiments}
We describe the experimental setups and discuss the performance of \model{} in terms of subjective as well as objective evaluations. Readers can listen to some audio samples on the demo webpage\footnote{\url{https://nvcnet.github.io/}}.

\subsection{Experimental setups} \label{sec:experiments:config}
All experiments are conducted on the VCTK data set~\citep{vctk_dataset}, which contains 44 hours of utterances from 109 speakers of English with various accents. Most sentences are extracted from newspapers and Rainbow Passage. In particular, each speaker reads a different set of sentences. We use a sampling rate of 22,050~Hz for all experiments. Utterances from the same speaker are randomly partitioned into training and test sets of ratio 9:1, respectively. \model{} is trained with the Adam optimizer using a learning rate of $10^{-4}$ with $\beta_1=0.5$ and $\beta_2=0.9$. We use a mini-batch size of 8 on 4 NVIDIA V100 GPUs. Training takes about 15 minutes per epoch. Inputs to \model{} are clips of length 32,768 samples ($\sim$1.5~seconds), which are randomly chosen from the utterances.

Following~\citet{Wester_2016}, we compute the naturalness and similarity scores of the converted utterances as subjective metrics. For the naturalness metric, a score ranging from totally unnatural (1) to totally natural (5) is assigned to each utterance. This metric takes into account the amount of distortion and artifacts presented in the utterances. For the similarity metric, a pair of utterances is presented in each test. This pair includes one converted utterance and one real utterance from the target speaker uttering the same sentence. Each subject is asked to assign a score ranging from different speakers (1) to the same speaker (5). The subject is explicitly instructed to listen beyond the distortion, misalignment, and mainly to focus on identifying the voice. The similarity metric aims to measure how well the converted utterance resembles the target speaker identity. There are in total 20 people that participated in our subjective evaluation.

We also consider a spoofing assessment as an objective metric. The spoofing measures the percentage of converted utterances being classified as the target speaker. We employ a spectrogram-based classifier trained on the same data split as for training the voice conversion system. A high spoofing value indicates that the voice conversion system can successfully convert the utterance to a target speaker, while a low value indicates that the speaker identity is not well captured by the system. Since we train the speaker identification classifier on clean data, the spoofing measure might be affected by distortions or artifacts. More details on our experimental setups are given in Appendix~\ref{appexdix:dataset}.

\subsection{Traditional voice conversion} \label{sec:experiments:many}
In this subsection, we compare \model{} with other state-of-the-art methods in non-parallel voice conversion, including Blow~\citep{NEURIPS2019_9426c311}, AutoVC~\citep{pmlr-v97-qian19c}, and StarGAN-VC2~\citep{Kaneko2019}. The details of these methods are given in Appendix~\ref{appendix:baseline}. To make a fair comparison, all methods use the same training and test sets. No extra data and transfer learning have been used in any of the competing methods. For AutoVC, we simply use the one-hot encoder for the speaker embeddings. Note that AutoVC uses the WaveNet~\cite{vandenOord_2016} vocoder pre-trained on the VCTK corpus, which may encode some information of the test utterances in the vocoder. This could be an unfair advantage. We also implement another version of our model, called \model{}$^{\dagger}$, which simply uses the one-hot encoder for the speaker embeddings. This allows us to study the importance of the speaker encoder.

\begin{figure}
    \centering
    \includegraphics[width=0.90\textwidth]{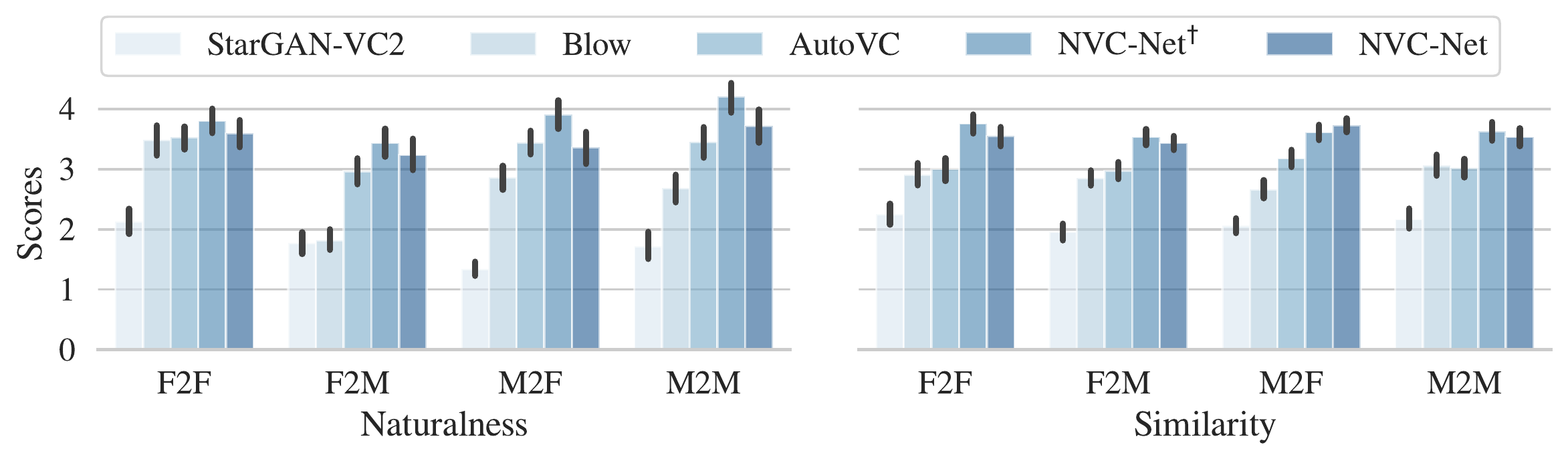}
    \caption{Subjective evaluation for traditional voice conversion settings with 95\% confidence intervals}
    \label{fig:mos}
    \vspace{-2mm}
\end{figure}

The subjective evaluation results are illustrated in Fig.~\ref{fig:mos}. Converted utterances are divided into four groups, including conversions of female to female (F2F), female to male (F2M), male to male (M2M), and male to female (M2F). As shown in the figure, \model{}$^{\dagger}$ and \model{} obtain good results over the four groups. Clearly, our methods show a big improvement over StarGAN-VC2 and Blow. In particular, \model{}$^{\dagger}$ achieves the best performance. Compared to AutoVC, which gives the current state-of-the-art results, \model{} yields very competitive performance. In terms of naturalness, \model{}$^{\dagger}$ yields slightly better than \model{}, while both perform equally well in terms of similarity. In addition, the spoofing results are shown in Table~\ref{table:evaluation}. The classifier used in this experiment achieves a high accuracy of 99.34\% on real speech. \model{}$^{\dagger}$ gives the highest spoofing score with 96.43\%, followed by \model{} with 93.66\%. Both \model{} and \model{}$^{\dagger}$ achieve superior performance than other competing methods with a clear margin.

\begin{table}
  \caption{Spoofing evaluations of the competing methods}
  \label{table:evaluation}
  \centering
  \begin{tabular}{lcccccc}
    \toprule
    Model & StarGAN-VC2 & AutoVC & Blow &  \model{}$^{\dagger}$ & \model{} \\
    \midrule
     Spoofing  & 19.08 & 82.46 &  89.39 & \textbf{96.43}  &  93.66\\ 
    \bottomrule
  \end{tabular}
  \vspace{-2mm}
\end{table}

\subsection{Zero-shot voice conversion}
\begin{figure}
    \centering
    \includegraphics[width=0.49\textwidth]{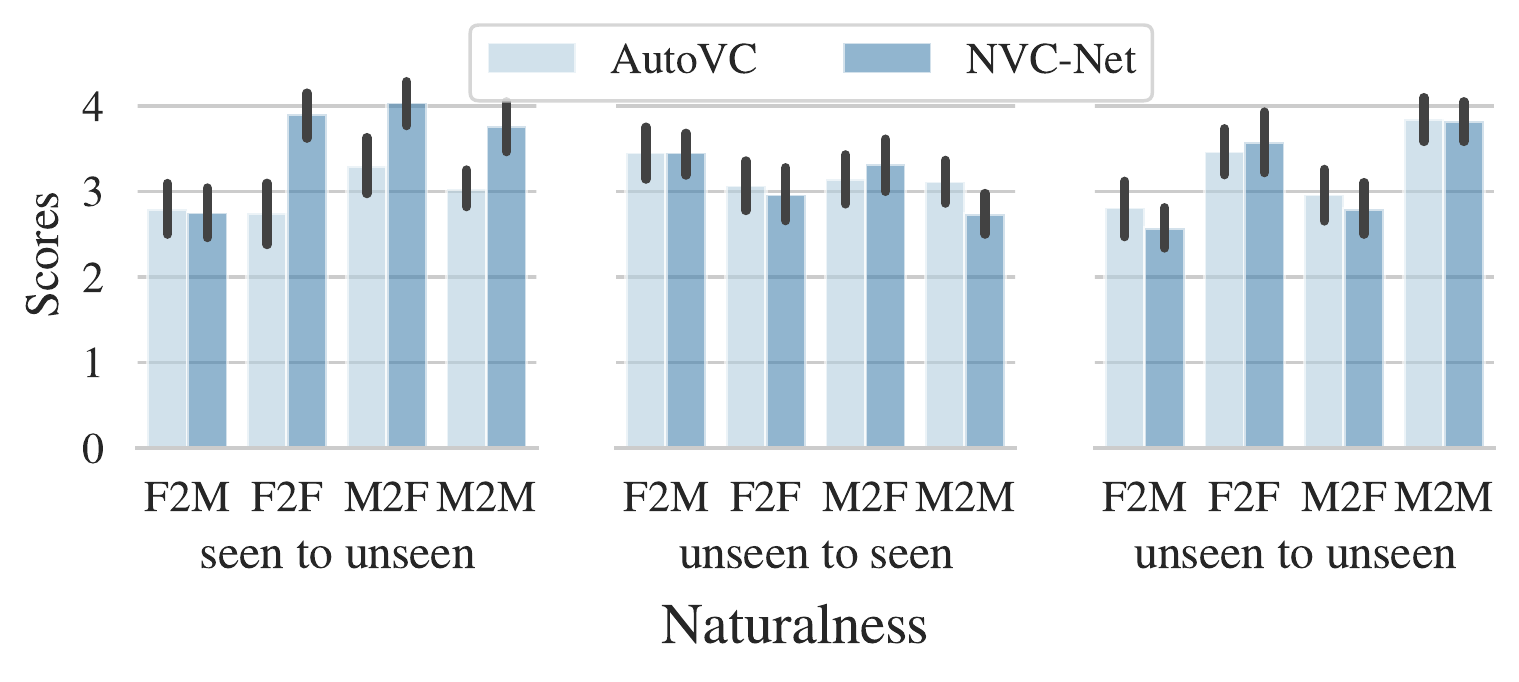}
    \includegraphics[width=0.49\textwidth]{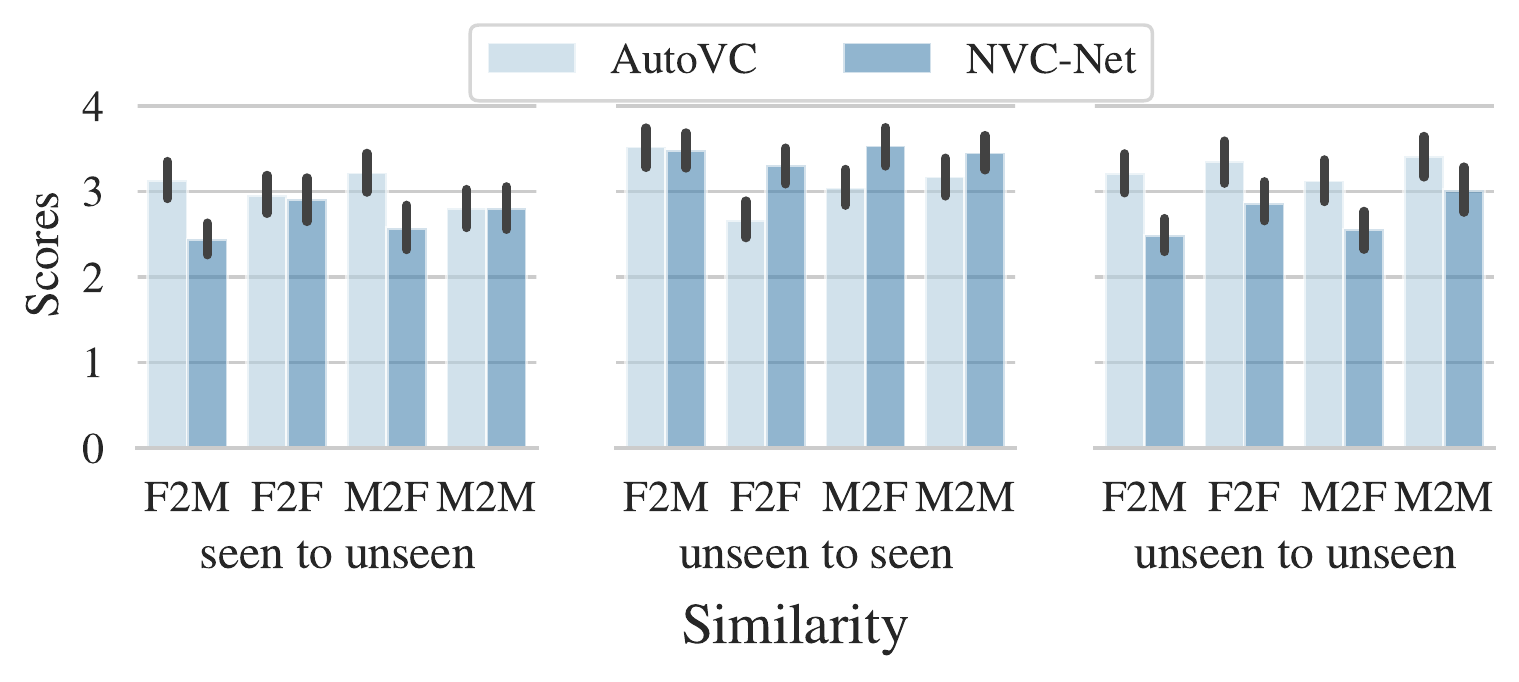}  
    \caption{Subjective evaluation for zero-shot voice conversion settings with 95\% confidence intervals}
    \label{fig:mos_unseen}
    \vspace{-2mm}
\end{figure}

In this subsection, we evaluate the generalization capability of \model{} on speakers that are unseen during training. Among the competing methods, only AutoVC supports zero-shot voice conversion. Therefore, we only report the results of \model{} against AutoVC. We conduct a similar experiment as in Subsection~\ref{sec:experiments:many}. In particular, the experiment is split into three different settings, including conversions of seen to unseen, unseen to seen, and unseen to unseen speakers. To get the best results for AutoVC, we use the pre-trained speaker encoder provided by the corresponding authors. The subjective evaluations are shown in Fig.~\ref{fig:mos_unseen}. Both \model{} and AutoVC achieve competitive results in terms of naturalness. AutoVC gives a slight improvement in terms of similarity. Note that the speaker encoder of AutoVC was trained on a large corpus of 3,549 speakers to make it generalizable to unseen speakers. Clearly, this gives an unfair advantage over our method in which no extra data are used. Further improvement for \model{} can be expected when more data are considered. However, we leave the evaluation of \model{} on a large-scale data set for future research because the main purpose here is to show that \model{} generalizes to unseen speakers without additional data.

\subsection{Further studies}
\textbf{Disentanglement analysis.} We perform an ablation study to verify the disentanglement of latent representation encoded by the  content and speaker encoders. The disentanglement is measured as the accuracy of a speaker identification classifier trained and validated on the latent representation. A high value indicates that the latent representation contains the speaker information. Note that with this VCTK data set, the content itself may reveal speaker identity as each speaker reads a different set of the  newspaper texts (split in the both training and test sets). For the speaker classifier, we simply use three fully-connected layers with 512 hidden neurons and ReLU activations. The results are shown in Table~\ref{table:disentanglement}. As expected, the classification accuracy is high when speaker embeddings are used and it substantially decreases when using content embeddings. The result confirms our hypothesis that the speaker encoder is able to disentangle the speaker identity from the content. 

\textbf{Inference speed and model size.}  Table~\ref{table:inference} shows the inference speed and model size for the competing methods. We did not report the parameters of the vocoders since AutoVC and StarGAN-VC2 can be used together with other choices of vocoders. Compared to the end-to-end Blow method, \model{} is significantly smaller, while being faster at inference time. Our model is capable of generating samples at a rate of 3661.65 kHz on an NVIDIA V100 GPU in full precision and 7.49 kHz on a single CPU core of Intel(R) Xeon(R) CPU E5-2695 v3 @ 2.30GHz. Gaining speed is mainly because \model{} is non-autoregressive and fully convolutional. This has an obvious advantage when used in real-time voice conversion systems. Interestingly, the inference speed of \model{} is 7x faster than the inference speed of WaveGlow as reported in~\cite{8683143}.

\begin{table}
  \parbox[t]{.3\linewidth}{
  \setlength{\tabcolsep}{1.5pt}
  \caption{Speaker identification accuracy (\%) using the content and speaker embeddings}
  \label{table:disentanglement}
  \centering
  \begin{tabular}{lrr}
    \toprule
     Model   & Content & Speaker \\
    \midrule
    \model{}$^{\dagger}$ & 19.21 & N/A \\
    \model{} & 24.15 & 99.22 \\
    \bottomrule
  \end{tabular}
  } \hfill
\parbox[t]{.65\linewidth}{
\setlength{\tabcolsep}{1.5pt}
  \caption{Model size and inference speed comparisons (\# of parameters of vocoders are not included in the methods with $^*$).}
  \label{table:inference}
  \centering
  \begin{tabular}{p{2.3cm}rrr}
    \toprule
    \multirow{2}{*}{Model}     & \# parameters      & Inference speed  & Inference speed\\
    &(in millions) & GPU (in kHz)  & CPU (in kHz)\\
    \midrule
    StarGAN-VC2$^{*}$ & \textbf{9.62} &   60.47    & \textbf{35.47} \\
    AutoVC$^{*}$ & 28.42  &    0.11 & 0.04 \\
    Blow     & 62.11 &  441.11 & 2.43\\
    \model{}    &  15.13 &  \textbf{3661.65} & 7.49 \\ 
    \bottomrule
  \end{tabular}} 
  \vspace{-2mm}
\end{table}

\begin{figure}
    \centering
    \begin{subfigure}[t]{.22\textwidth}
      \centering
      \includegraphics[width=\linewidth]{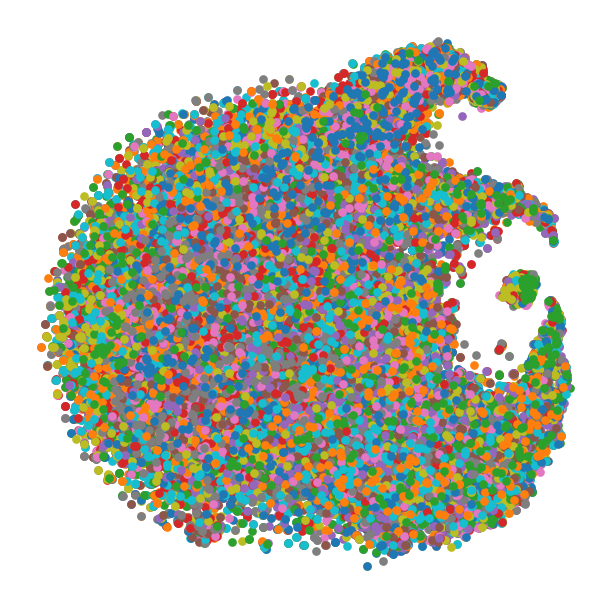}  
      \caption{Seen}
      \label{fig:content}
    \end{subfigure}
    \hfill
    \begin{subfigure}[t]{.22\textwidth}
      \centering
      \includegraphics[width=\linewidth]{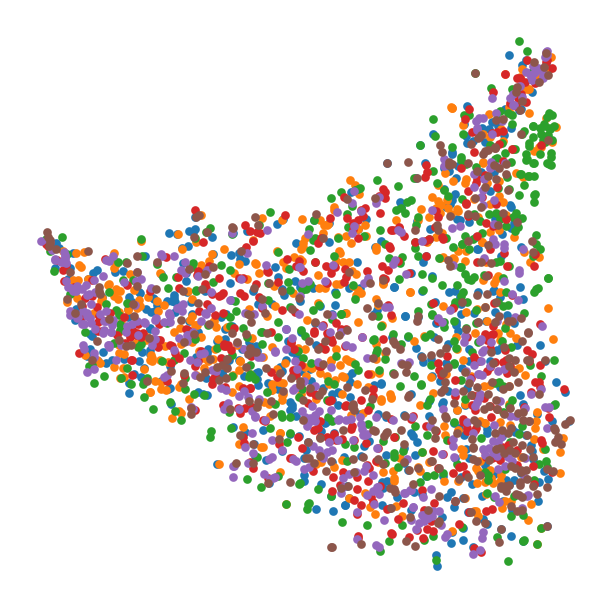}  
      \caption{Unseen}
      \label{fig:content}
    \end{subfigure}
    \hfill
    \begin{subfigure}[t]{0.22\textwidth}
      \centering
      \includegraphics[width=\linewidth]{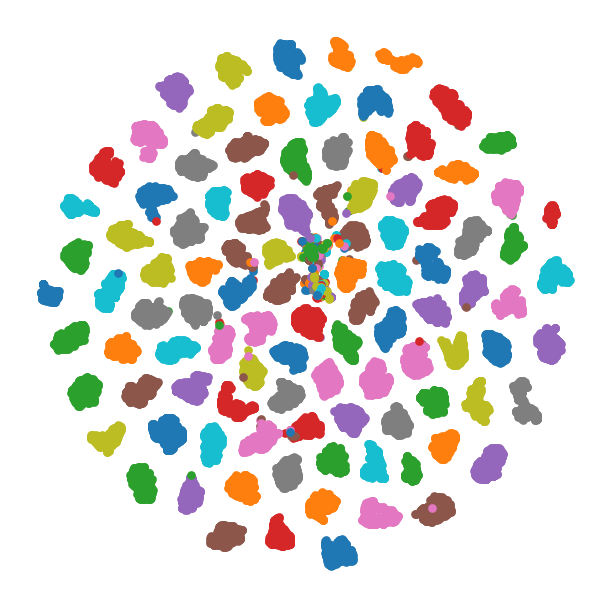}  
      \caption{Seen}
      \label{fig:sub-second}
    \end{subfigure}
    \hfill
    \begin{subfigure}[t]{0.22\textwidth}
      \centering
      \includegraphics[width=\linewidth]{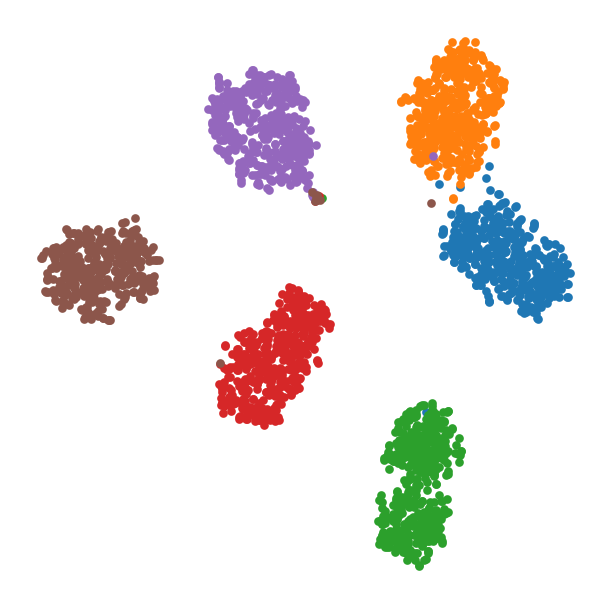}  
      \caption{Unseen}
      \label{fig:sub-second}
    \end{subfigure}
    \caption{The Barnes-Hut t-SNE visualization~\cite{JMLR:v15:vandermaaten14a} of the content embeddings (see (a) and (b)) and speaker embeddings (see (c) and (d)) of utterances from seen and unseen speakers.} 
    \label{figure:tsne}
    \vspace{-2mm}
\end{figure}

\textbf{Latent embedding visualization.} Figure~\ref{figure:tsne} illustrates the content and speaker embeddings in 2D using the Barnes-Hut t-SNE visualization~\cite{JMLR:v15:vandermaaten14a}. Based on speaker embeddings, utterances of the same speaker are clustered together, while those of different speakers are well separated. It is important to note that we do not directly impose any constraints on these speaker embeddings. Due to the KL divergence regularization, the speaker latent space is smooth, allowing to sample a speaker embedding from a prior Gaussian distribution. On the other hand, based on content embeddings, utterances are scattered over the whole space. Content embeddings of the same speaker are not nicely clustered together. Our results indicate that speaker information is not embedded in the content representation.

\section{Conclusions} \label{sec:conclusions}
In this paper, we have introduced \model{}, an adversarial neural network which is trained in an end-to-end manner for voice conversion. Rather than constraining the model to work with typical intermediate representation (\textit{e.g.}, spectrograms, linguistic features, etc), \model{} is able to exploit a good internal representation through a combination of adversarial feedback and reconstruction loss. In particular, \model{} aims to disentangle the speaker identity from the linguistic content. Our empirical studies have confirmed that \model{} yields very competitive results in traditional voice conversion settings as well as in zero-shot voice conversion settings. Compared to other voice conversion methods, \model{} is very efficient at inference. Although \model{} can synthesize diverse samples, it still lacks of explicit controls over other aspects of speech, \textit{e.g.}, rhythm and prosody. In future work, we will focus on addressing this issue. Another promising direction is to extend \model{} to other speech conversion tasks such as music and singing voice.

\section*{Broad impact}
Voice conversion and, more generally, speech synthesis models have a wide range of applications with various impact on the society. 
In the entertainment industry, automatic content generation will likely spur new types of personalised games with scenes generated along the way. 
It has also the potential to make dubbing easier and cheaper allowing communities to access a vast amount of content previously unavailable in their languages. In the medical domain, the model presented in this paper can lead to new types of communication aids for the speech impaired. In particular, we envision communication aids that preserve the original voice of a person who has lost the ability to speak. 
Amongst the applications with positive societal impact, voice conversion can be used for speaker de-identification which helps the preservation of privacy. On the other hand, it can also be used for generating deepfakes and contribute to the propagation of `fake news'. For this reason, we believe that it is critical for the research community to also conduct research on the detection of manipulated images~\cite{lyu2020deepfake, rossler2019faceforensics} and audio~\cite{wang2020deepsonar, albadawy2019detecting}. 
Improving the computational efficiency of voice conversion models has motivated the work presented in this paper. In particular, the proposed end-to-end model does not require the use of vocoder at inference which typically amounts for most of the computation. Subjective evaluations show that our model can achieve competitive results with the state-of-the-art AutoVC (see Figs.~\ref{fig:mos} and~\ref{fig:mos_unseen}) while being four orders of magnitude faster on GPU (see Table~\ref{table:inference}). These results suggest that the voice conversion inference can be performed massively with less hardware and less energy consumption. In general, we believe that the energy consumption and environmental impact should be an important criteria when developing novel models. While we are all excited and impressed by the performance of gigantic models, the research community should also consider weighing the positive societal impact of such models against the environmental impact of using them. We refer the readers to the work from~\citet{strubell2019energy} for further discussions on this aspect. Finally, the results in Table~\ref{table:inference} show that we are also approaching a real-time inference on a single CPU core without any particular code optimization. This suggests that voice conversion applications will be able to be executed on local devices, helping to preserve the anonymity and privacy of the user. 

\section*{Acknowledgments}
The authors would like to thank Stefan Uhlich, Ali Aroudi, Marc Ferras Font, Lukas Mauch and other colleagues at Sony Europe B.V., Stuttgart Laboratory 1 for many valuable discussions and suggestions.

\bibliographystyle{abbrvnat}

{
\small

\begin{thebibliography}{52}
\providecommand{\natexlab}[1]{#1}
\providecommand{\url}[1]{\texttt{#1}}
\expandafter\ifx\csname urlstyle\endcsname\relax
  \providecommand{\doi}[1]{doi: #1}\else
  \providecommand{\doi}{doi: \begingroup \urlstyle{rm}\Url}\fi

\bibitem[AlBadawy et~al.(2019)AlBadawy, Lyu, and Farid]{albadawy2019detecting}
E.~A. AlBadawy, S.~Lyu, and H.~Farid.
\newblock Detecting ai-synthesized speech using bispectral analysis.
\newblock In \emph{Proceedings of the Conference on Computer Vision and Pattern
  Recognition Workshops}, pages 104--109, 2019.

\bibitem[Bi{\'n}kowski et~al.(2020)Bi{\'n}kowski, Donahue, Dieleman, Clark,
  Elsen, Casagrande, Cobo, and Simonyan]{binkowski2019high}
M.~Bi{\'n}kowski, J.~Donahue, S.~Dieleman, A.~Clark, E.~Elsen, N.~Casagrande,
  L.~C. Cobo, and K.~Simonyan.
\newblock High fidelity speech synthesis with adversarial networks.
\newblock In \emph{Proceedings of the International Conference on Learning
  Representations}, 2020.

\bibitem[chieh Chou and Lee(2019)]{Chou2019}
J.~chieh Chou and H.-Y. Lee.
\newblock One-shot voice conversion by separating speaker and content
  representations with instance normalization.
\newblock In \emph{Proceedings of the Conference of the International Speech
  Communication Association}, pages 664--668, 2019.

\bibitem[Choi et~al.(2018)Choi, Choi, Kim, Ha, Kim, and Choo]{Choi_2018_CVPR}
Y.~Choi, M.~Choi, M.~Kim, J.-W. Ha, S.~Kim, and J.~Choo.
\newblock Stargan: Unified generative adversarial networks for multi-domain
  image-to-image translation.
\newblock In \emph{Proceedings of the Conference on Computer Vision and Pattern
  Recognition}, pages 8789--8797, 2018.

\bibitem[Donahue et~al.(2019)Donahue, McAuley, and
  Puckette]{donahue2018adversarial}
C.~Donahue, J.~McAuley, and M.~Puckette.
\newblock Adversarial audio synthesis.
\newblock In \emph{Proceedings of the International Conference on Learning
  Representations}, 2019.

\bibitem[Dumoulin et~al.(2017)Dumoulin, Shlens, and
  Kudlur]{dumoulin2017learned}
V.~Dumoulin, J.~Shlens, and M.~Kudlur.
\newblock A learned representation for artistic style.
\newblock In \emph{Proceedings of the International Conference on Learning
  Representations}, 2017.

\bibitem[Engel et~al.(2017)Engel, Resnick, Roberts, Dieleman, Norouzi, Eck, and
  Simonyan]{pmlr-v70-engel17a}
J.~Engel, C.~Resnick, A.~Roberts, S.~Dieleman, M.~Norouzi, D.~Eck, and
  K.~Simonyan.
\newblock Neural audio synthesis of musical notes with {W}ave{N}et
  autoencoders.
\newblock In \emph{Proceedings of the International Conference on Machine
  Learning}, pages 1068--1077, 2017.

\bibitem[Engel et~al.(2020)Engel, Hantrakul, Gu, and Roberts]{engel2020ddsp}
J.~Engel, L.~Hantrakul, C.~Gu, and A.~Roberts.
\newblock {DDSP}: Differentiable digital signal processing.
\newblock In \emph{Proceedings of the International Conference on Learning
  Representations}, 2020.

\bibitem[{Fang} et~al.(2018){Fang}, {Yamagishi}, {Echizen}, and
  {Lorenzo-Trueba}]{8462342}
F.~{Fang}, J.~{Yamagishi}, I.~{Echizen}, and J.~{Lorenzo-Trueba}.
\newblock High-quality nonparallel voice conversion based on cycle-consistent
  adversarial network.
\newblock In \emph{Proceedings of the International Conference on Acoustics,
  Speech and Signal Processing}, pages 5279--5283, 2018.

\bibitem[Goodfellow et~al.(2014)Goodfellow, Pouget-Abadie, Mirza, Xu,
  Warde-Farley, Ozair, Courville, and Bengio]{NIPS2014_5ca3e9b1}
I.~Goodfellow, J.~Pouget-Abadie, M.~Mirza, B.~Xu, D.~Warde-Farley, S.~Ozair,
  A.~Courville, and Y.~Bengio.
\newblock Generative adversarial nets.
\newblock In \emph{Advances in Neural Information Processing Systems}, pages
  2672--2680, 2014.

\bibitem[Helander et~al.(2008)Helander, Schwarz, Nurminen, Silen, and
  Gabbouj]{helander2008impact}
E.~Helander, J.~Schwarz, J.~Nurminen, H.~Silen, and M.~Gabbouj.
\newblock On the impact of alignment on voice conversion performance.
\newblock In \emph{Proceedings of the Conference of the International Speech
  Communication Association}, pages 1453--1456, 2008.

\bibitem[Hendrycks and Gimpel(2016)]{hendrycks2016gaussian}
D.~Hendrycks and K.~Gimpel.
\newblock Gaussian error linear units (gelus).
\newblock \emph{arXiv preprint arXiv:1606.08415}, 2016.

\bibitem[{Hsu} et~al.(2016){Hsu}, {Hwang}, {Wu}, {Tsao}, and {Wang}]{7820786}
C.~{Hsu}, H.~{Hwang}, Y.~{Wu}, Y.~{Tsao}, and H.~{Wang}.
\newblock Voice conversion from non-parallel corpora using variational
  auto-encoder.
\newblock In \emph{Proceedings of the Asia-Pacific Signal and Information
  Processing Association Annual Summit and Conference}, pages 1--6, 2016.

\bibitem[Hsu et~al.(2017)Hsu, Hwang, Wu, Tsao, and Wang]{hsu2017voice}
C.-C. Hsu, H.-T. Hwang, Y.-C. Wu, Y.~Tsao, and H.-M. Wang.
\newblock Voice conversion from unaligned corpora using variational
  autoencoding wasserstein generative adversarial networks.
\newblock In \emph{Proceedings of the Conference of the International Speech
  Communication Association}, pages 3364--3368, 2017.

\bibitem[{Isola} et~al.(2017){Isola}, {Zhu}, {Zhou}, and {Efros}]{8100115}
P.~{Isola}, J.~{Zhu}, T.~{Zhou}, and A.~A. {Efros}.
\newblock Image-to-image translation with conditional adversarial networks.
\newblock In \emph{Proceedings of the Conference on Computer Vision and Pattern
  Recognition}, pages 5967--5976, 2017.

\bibitem[Kalchbrenner et~al.(2018)Kalchbrenner, Elsen, Simonyan, Noury,
  Casagrande, Lockhart, Stimberg, van~den Oord, Dieleman, and
  Kavukcuoglu]{pmlr-v80-kalchbrenner18a}
N.~Kalchbrenner, E.~Elsen, K.~Simonyan, S.~Noury, N.~Casagrande, E.~Lockhart,
  F.~Stimberg, A.~van~den Oord, S.~Dieleman, and K.~Kavukcuoglu.
\newblock Efficient neural audio synthesis.
\newblock In \emph{Proceedings of the International Conference on Machine
  Learning}, pages 2410--2419, 2018.

\bibitem[{Kameoka} et~al.(2018){Kameoka}, {Kaneko}, {Tanaka}, and
  {Hojo}]{8639535}
H.~{Kameoka}, T.~{Kaneko}, K.~{Tanaka}, and N.~{Hojo}.
\newblock {StarGAN-VC}: Non-parallel many-to-many voice conversion using star
  generative adversarial networks.
\newblock In \emph{Proceedings of the Spoken Language Technology Workshop},
  pages 266--273, 2018.

\bibitem[{Kameoka} et~al.(2019){Kameoka}, {Kaneko}, {Tanaka}, and
  {Hojo}]{8718381}
H.~{Kameoka}, T.~{Kaneko}, K.~{Tanaka}, and N.~{Hojo}.
\newblock {ACVAE-VC}: Non-parallel voice conversion with auxiliary classifier
  variational autoencoder.
\newblock \emph{IEEE/ACM Transactions on Audio, Speech, and Language
  Processing}, 27:\penalty0 1432--1443, 2019.

\bibitem[Kaneko and Kameoka(2017)]{kaneko2017parallel}
T.~Kaneko and H.~Kameoka.
\newblock Parallel-data-free voice conversion using cycle-consistent
  adversarial networks.
\newblock \emph{arXiv preprint arXiv:1711.11293}, 2017.

\bibitem[{Kaneko} and {Kameoka}(2018)]{8553236}
T.~{Kaneko} and H.~{Kameoka}.
\newblock {CycleGAN-VC}: Non-parallel voice conversion using cycle-consistent
  adversarial networks.
\newblock In \emph{Proceedings of the European Signal Processing Conference},
  pages 2100--2104, 2018.

\bibitem[Kaneko et~al.(2017)Kaneko, Kameoka, Hiramatsu, and
  Kashino]{Kaneko2017}
T.~Kaneko, H.~Kameoka, K.~Hiramatsu, and K.~Kashino.
\newblock Sequence-to-sequence voice conversion with similarity metric learned
  using generative adversarial networks.
\newblock In \emph{Proceedings of the Conference of the International Speech
  Communication Association}, pages 1283--1287, 2017.

\bibitem[Kaneko et~al.(2019)Kaneko, Kameoka, Tanaka, and Hojo]{Kaneko2019}
T.~Kaneko, H.~Kameoka, K.~Tanaka, and N.~Hojo.
\newblock {StarGAN-VC2}: Rethinking conditional methods for stargan-based voice
  conversion.
\newblock In \emph{Proceedings of the Conference of the International Speech
  Communication Association}, pages 679--683, 2019.

\bibitem[Karras et~al.(2020)Karras, Laine, Aittala, Hellsten, Lehtinen, and
  Aila]{Karras_2020_CVPR}
T.~Karras, S.~Laine, M.~Aittala, J.~Hellsten, J.~Lehtinen, and T.~Aila.
\newblock Analyzing and improving the image quality of stylegan.
\newblock In \emph{Proceedings of the Conference on Computer Vision and Pattern
  Recognition}, pages 8110--8119, 2020.

\bibitem[Kawahara et~al.(1999)Kawahara, Masuda-Katsuse, and
  de~Cheveign\'{e}]{10.1016/S0167}
H.~Kawahara, I.~Masuda-Katsuse, and A.~de~Cheveign\'{e}.
\newblock Restructuring speech representations using a pitch-adaptive
  time-frequency smoothing and an instantaneous-frequency-based f0 extraction:
  Possible role of a repetitive structure in sounds.
\newblock \emph{Speech Communication}, 27:\penalty0 187--207, 1999.

\bibitem[Kingma and Dhariwal(2018)]{NEURIPS2018_d139db6a}
D.~P. Kingma and P.~Dhariwal.
\newblock Glow: Generative flow with invertible 1x1 convolutions.
\newblock In \emph{Advances in Neural Information Processing Systems}, pages
  10236--10245, 2018.

\bibitem[Kingma and Welling(2014)]{kingma:vae}
D.~P. Kingma and M.~Welling.
\newblock Auto-encoding variational {Bayes}.
\newblock In \emph{Proceedings of the International Conference on Learning
  Representations}, 2014.

\bibitem[Kumar et~al.(2019)Kumar, Kumar, de~Boissiere, Gestin, Teoh, Sotelo,
  de~Br\'{e}bisson, Bengio, and Courville]{NEURIPS2019_6804c9bc}
K.~Kumar, R.~Kumar, T.~de~Boissiere, L.~Gestin, W.~Z. Teoh, J.~Sotelo,
  A.~de~Br\'{e}bisson, Y.~Bengio, and A.~C. Courville.
\newblock {MelGAN}: Generative adversarial networks for conditional waveform
  synthesis.
\newblock In \emph{Advances in Neural Information Processing Systems}, pages
  14910--14921, 2019.

\bibitem[Larsen et~al.(2016)Larsen, Sønderby, Larochelle, and
  Winther]{pmlr-v48-larsen16}
A.~B.~L. Larsen, S.~K. Sønderby, H.~Larochelle, and O.~Winther.
\newblock Autoencoding beyond pixels using a learned similarity metric.
\newblock In \emph{Proceedings of the International Conference on Machine
  Learning}, pages 1558--1566, 2016.

\bibitem[Liu et~al.(2018)Liu, Ling, Jiang, Zhou, and Dai]{Liu2018}
L.-J. Liu, Z.-H. Ling, Y.~Jiang, M.~Zhou, and L.-R. Dai.
\newblock {WaveNet} vocoder with limited training data for voice conversion.
\newblock In \emph{Proceedings of the Conference of the International Speech
  Communication Association}, pages 1983--1987, 2018.

\bibitem[Liu et~al.(2019)Liu, Huang, Mallya, Karras, Aila, Lehtinen, and
  Kautz]{05cf1}
M.-Y. Liu, X.~Huang, A.~Mallya, T.~Karras, T.~Aila, J.~Lehtinen, and J.~Kautz.
\newblock Few-shot unsupervised image-to-image translation.
\newblock In \emph{Proceedings of the International Conference on Computer
  Vision}, 2019.

\bibitem[Lorenzo-Trueba et~al.(2018)Lorenzo-Trueba, Yamagishi, Toda, Saito,
  Villavicencio, Kinnunen, and Ling]{Lorenzo-Trueba2018}
J.~Lorenzo-Trueba, J.~Yamagishi, T.~Toda, D.~Saito, F.~Villavicencio,
  T.~Kinnunen, and Z.~Ling.
\newblock The voice conversion challenge 2018: Promoting development of
  parallel and nonparallel methods.
\newblock In \emph{Proceedings of the Speaker and Language Recognition
  Workshop}, pages 195--202, 2018.

\bibitem[Lyu(2020)]{lyu2020deepfake}
S.~Lyu.
\newblock Deepfake detection: Current challenges and next steps.
\newblock In \emph{Proceedings of the International Conference on Multimedia \&
  Expo Workshops}, pages 1--6, 2020.

\bibitem[Mao et~al.(2017)Mao, Li, Xie, Lau, Wang, and Smolley]{8237566}
X.~Mao, Q.~Li, H.~Xie, R.~Y. Lau, Z.~Wang, and S.~P. Smolley.
\newblock Least squares generative adversarial networks.
\newblock In \emph{Proceedings of the International Conference on Computer
  Vision}, pages 2813--2821, 2017.

\bibitem[Morise et~al.(2016)Morise, Yokomori, and Ozawa]{Masanori}
M.~Morise, F.~Yokomori, and K.~Ozawa.
\newblock {WORLD}: A vocoder-based high-quality speech synthesis system for
  real-time applications.
\newblock \emph{IEICE Transactions on Information and Systems}, E99-D:\penalty0
  1877--1884, 2016.

\bibitem[Nakamura et~al.(2012)Nakamura, Toda, Saruwatari, and
  Shikano]{NAKAMURA2012134}
K.~Nakamura, T.~Toda, H.~Saruwatari, and K.~Shikano.
\newblock Speaking-aid systems using gmm-based voice conversion for
  electrolaryngeal speech.
\newblock \emph{Speech Communication}, 54:\penalty0 134--146, 2012.

\bibitem[{Prenger} et~al.(2019){Prenger}, {Valle}, and {Catanzaro}]{8683143}
R.~{Prenger}, R.~{Valle}, and B.~{Catanzaro}.
\newblock {Waveglow}: A flow-based generative network for speech synthesis.
\newblock In \emph{Proceedings of the International Conference on Acoustics,
  Speech and Signal Processing}, pages 3617--3621, 2019.

\bibitem[Qian et~al.(2019)Qian, Zhang, Chang, Yang, and
  Hasegawa-Johnson]{pmlr-v97-qian19c}
K.~Qian, Y.~Zhang, S.~Chang, X.~Yang, and M.~Hasegawa-Johnson.
\newblock {A}uto{VC}: Zero-shot voice style transfer with only autoencoder
  loss.
\newblock In \emph{Proceedings of the International Conference on Machine
  Learning}, pages 5210--5219, 2019.

\bibitem[Rossler et~al.(2019)Rossler, Cozzolino, Verdoliva, Riess, Thies, and
  Nie{\ss}ner]{rossler2019faceforensics}
A.~Rossler, D.~Cozzolino, L.~Verdoliva, C.~Riess, J.~Thies, and M.~Nie{\ss}ner.
\newblock Faceforensics++: Learning to detect manipulated facial images.
\newblock In \emph{Proceedings of the International Conference on Computer
  Vision}, pages 1--11, 2019.

\bibitem[Salimans and Kingma(2016)]{10.5555/3157096.3157197}
T.~Salimans and D.~P. Kingma.
\newblock Weight normalization: A simple reparameterization to accelerate
  training of deep neural networks.
\newblock In \emph{Advances in Neural Information Processing Systems}, pages
  901--909, 2016.

\bibitem[Serr\`{a} et~al.(2019)Serr\`{a}, Pascual, and
  Segura~Perales]{NEURIPS2019_9426c311}
J.~Serr\`{a}, S.~Pascual, and C.~Segura~Perales.
\newblock Blow: a single-scale hyperconditioned flow for non-parallel raw-audio
  voice conversion.
\newblock In \emph{Advances in Neural Information Processing Systems}, pages
  6793--6803, 2019.

\bibitem[{Sisman} et~al.(2021){Sisman}, {Yamagishi}, {King}, and {Li}]{9262021}
B.~{Sisman}, J.~{Yamagishi}, S.~{King}, and H.~{Li}.
\newblock An overview of voice conversion and its challenges: From statistical
  modeling to deep learning.
\newblock \emph{IEEE/ACM Transactions on Audio, Speech, and Language
  Processing}, 29:\penalty0 132--157, 2021.

\bibitem[Strubell et~al.(2019)Strubell, Ganesh, and
  McCallum]{strubell2019energy}
E.~Strubell, A.~Ganesh, and A.~McCallum.
\newblock Energy and policy considerations for deep learning in {NLP}.
\newblock In \emph{Proceedings of the Annual Meeting of the Association for
  Computational Linguistics}, pages 3645--3650, 2019.

\bibitem[{Tao} et~al.(2010){Tao}, {Zhang}, {Nurminen}, {Tian}, and
  {Wang}]{5485203}
J.~{Tao}, M.~{Zhang}, J.~{Nurminen}, J.~{Tian}, and X.~{Wang}.
\newblock Supervisory data alignment for text-independent voice conversion.
\newblock \emph{IEEE Transactions on Audio, Speech, and Language Processing},
  18:\penalty0 932--943, 2010.

\bibitem[van~den Oord et~al.(2016)van~den Oord, Dieleman, Zen, Simonyan,
  Vinyals, Graves, Kalchbrenner, Senior, and Kavukcuoglu]{vandenOord_2016}
A.~van~den Oord, S.~Dieleman, H.~Zen, K.~Simonyan, O.~Vinyals, A.~Graves,
  N.~Kalchbrenner, A.~Senior, and K.~Kavukcuoglu.
\newblock {WaveNet}: A generative model for raw audio.
\newblock In \emph{Proceedings of the International Speech Communication
  Association Workshop}, pages 125--125, 2016.

\bibitem[van~der Maaten(2014)]{JMLR:v15:vandermaaten14a}
L.~van~der Maaten.
\newblock Accelerating t-{SNE} using tree-based algorithms.
\newblock \emph{The Journal of Machine Learning Research}, 15:\penalty0
  3221--3245, 2014.

\bibitem[Veaux et~al.(2017)Veaux, Yamagishi, and MacDonald]{vctk_dataset}
C.~Veaux, J.~Yamagishi, and K.~MacDonald.
\newblock Superseded - cstr vctk corpus: English multi-speaker corpus for cstr
  voice cloning toolkit, 2017.
\newblock URL \url{http://datashare.is.ed.ac.uk/handle/10283/2651}.

\bibitem[Wang et~al.(2020)Wang, Juefei-Xu, Huang, Guo, Xie, Ma, and
  Liu]{wang2020deepsonar}
R.~Wang, F.~Juefei-Xu, Y.~Huang, Q.~Guo, X.~Xie, L.~Ma, and Y.~Liu.
\newblock Deepsonar: Towards effective and robust detection of ai-synthesized
  fake voices.
\newblock In \emph{Proceedings of the ACM International Conference on
  Multimedia}, pages 1207--1216, 2020.

\bibitem[Wester et~al.(2016)Wester, Wu, and Yamagishi]{Wester_2016}
M.~Wester, Z.~Wu, and J.~Yamagishi.
\newblock Analysis of the voice conversion challenge 2016 evaluation results.
\newblock In \emph{Proceedings of the Conference of the International Speech
  Communication Association}, pages 1637--1641, 2016.

\bibitem[Wu and Lee(2020)]{9053854}
D.-Y. Wu and H.-y. Lee.
\newblock One-shot voice conversion by vector quantization.
\newblock In \emph{Proceedings of the International Conference on Acoustics,
  Speech and Signal Processing}, pages 7734--7738, 2020.

\bibitem[Wu et~al.(2018)Wu, Kobayashi, Hayashi, Tobing, and Toda]{Wu2018}
Y.-C. Wu, K.~Kobayashi, T.~Hayashi, P.~L. Tobing, and T.~Toda.
\newblock Collapsed speech segment detection and suppression for {WaveNet}
  vocoder.
\newblock In \emph{Proceedings of the Conference of the International Speech
  Communication Association}, pages 1988--1992, 2018.

\bibitem[Yamamoto et~al.(2020)Yamamoto, Song, and Kim]{9053795}
R.~Yamamoto, E.~Song, and J.-M. Kim.
\newblock Parallel wavegan: A fast waveform generation model based on
  generative adversarial networks with multi-resolution spectrogram.
\newblock In \emph{Proceedings of the International Conference on Acoustics,
  Speech and Signal Processing}, pages 6199--6203, 2020.

\bibitem[Zhang et~al.(2020)Zhang, Liu, Chen, Hu, Jiang, Ling, and
  Dai]{Zhang2020}
J.-X. Zhang, L.-J. Liu, Y.-N. Chen, Y.-J. Hu, Y.~Jiang, Z.-H. Ling, and L.-R.
  Dai.
\newblock Voice conversion by cascading automatic speech recognition and
  text-to-speech synthesis with prosody transfer.
\newblock In \emph{Proceedings of the Joint Workshop for the Blizzard Challenge
  and Voice Conversion Challenge}, pages 121--125, 2020.

\end{thebibliography}
}

\newpage
\appendix

\section{Appendix}
\subsection{NVC-Net from the probabilistic model perspective} \label{appendix:loss}
In this section, we provide a more formal explanation of using the Kullback–Leibler divergence as a regularizer in Eq.~(\ref{eq:kl}). Our derivation follows the arguments of the standard VAE~\cite{kingma:vae}. We assume that the observed utterance $\vec{x}$ is generated from two independent latent variables $\vec{c}$ and $\vec{z}$, corresponding to the speech content and speaker identity of $\vec{x}$, respectively. Let $q(\vec{c}|\vec{x})$ and $q(\vec{z}|\vec{x})$ denote approximate posteriors for the latent variables and we assume that $q(\vec{c}, \vec{z}|\vec{x}) = q(\vec{c}|\vec{x}) q(\vec{z}|\vec{x})$. Because our content encoder is deterministic, this implies $q(\vec{c}|\vec{x})=1$. We follow the variational principle
\begin{align}
   \log p(\vec{x}) &= \log \iint p(\vec{x}, \vec{c}, \vec{z}) d\vec{c} d\vec{z} \notag\\
    &= \log \iint \frac{q(\vec{c},\vec{z}|\vec{x})}{q(\vec{c},\vec{z}|\vec{x})} p(\vec{x} | \vec{c}, \vec{z}) p(\vec{c}, \vec{z}) d\vec{c} d\vec{z} \notag \\
    &= \log \iint q(\vec{c},\vec{z}|\vec{x}) \frac{p(\vec{c}, \vec{z})}{q(\vec{c},\vec{z}|\vec{x})} p(\vec{x} | \vec{c}, \vec{z})   d\vec{c} d\vec{z} \notag\\
    & \ge \iint q(\vec{c}|\vec{x})  q(\vec{z}|\vec{x}) \log \frac{p(\vec{z})p(\vec{c})}{q(\vec{c}|\vec{x})  q(\vec{z}|\vec{x})} d\vec{c}  d\vec{z}  + \iint q(\vec{c},\vec{z}|\vec{x}) \log p(\vec{x}|\vec{c}, \vec{z}) d\vec{c} d\vec{z} \notag\\
    &= \int q(\vec{z} | \vec{x}) \log \frac{p(\vec{z})}{q(\vec{z}|\vec{x})} d\vec{z} + \iint q(\vec{z}|\vec{x}) \log p(\vec{c}) d\vec{c} d\vec{z} + \EX_{q(\vec{c}, \vec{z}|\vec{x})}[\log p(\vec{x}|\vec{c}, \vec{z})] \notag\\
    &= -\mathbb{D}_{\text{KL}}(q(\vec{z}|\vec{x})\| p(\vec{z})) + \int \log p(\vec{c}) d\vec{c} + \EX_{q(\vec{c}, \vec{z}|\vec{x})}[\log p(\vec{x}|\vec{c}, \vec{z})]\,. \label{eq:dev}
\end{align}
We choose the prior distributions $p(\vec{z})$ to be a zero-mean unit-variance Gaussian  and $p(\vec{c})$ to be a uniform distribution on the surface of unit sphere, \textit{i.e.},
\begin{align*}
    \vec{z}&\sim \mathcal{N}(\vec{z}|\vec{0}, \vec{I})\,,\\
    \vec{c} &\sim \mathcal{U}(\vec{c}| (\textstyle \sum_j c^2_{ij})^{1/2} =1)\,.
\end{align*}
Therefore, the second term in Eq.~(\ref{eq:dev}) is a constant and equal to $-L_{\text{con}}d_{\text{con}}\log(A)$, where $L_{\text{con}}$, $d_{\text{con}}$ are the temporal and spatial dimensions of $\vec{c}$, respectively, and $A$ is the surface area of a $d_{\text{con}}$-dimensional unit ball. Equation~(\ref{eq:dev}) is also known as the evidence lower bound (ELBO) or the negative variational free energy. One can see that maximizing the ELBO is equivalent to minimize $\mathbb{D}_{\text{KL}}(q(\vec{z}|\vec{x})\| p(\vec{z}))$ and to maximize $\EX_{q(\vec{c}, \vec{z}|\vec{x})}[\log p(\vec{x}|\vec{c}, \vec{z})]$.  The former term corresponds to our Kullback–Leibler regularization in Eq.~(\ref{eq:kl}) and the latter term corresponds to our reconstruction loss in Eq.~(\ref{eq:recon}). In other words, the loss function of \model{} implicitly performs the maximum likelihood estimation to learn the parameters.

\subsection{Model architectures} \label{appendix:network}
In this section, we describe in more detail the architecture design for \model{}. In order to stabilize the training and to improve the convergence of GANs, weight normalization~\cite{10.5555/3157096.3157197} is employed in all layers of \model{}, including encoders, generator, and discriminators. We do not use batch normalization or instance normalization as they tend to remove important information of speech~\cite{NEURIPS2019_6804c9bc}. 

Let $T$ denote the temporal length of the input audio waveform and $d_{\text{con}}$ denote the spatial dimension of the content embedding. The network architecture of the content encoder is shown in Table~\ref{tab:content_encoder}. The temporal resolution of input is reduced 256 times, which is done in 4 stages of 2x, 2x, 8x, and 8x downsampling. A residual block (ResidualBlock) consists of 4 layers of dilated convolutions with dilation 1, 3, 9, and 27, inducing a receptive field of 81. In our experiments, we set $d_{\text{con}}=4$.

We use 80-band mel-spectrograms as inputs for the speaker encoder, where the FFT, window, and hop size are set to 1024, 1024, and 256, respectively. Let $M$ denote the temporal length of the mel-spectrogram.  Table~\ref{tab:speaker_encoder} shows the network architecture of the speaker encoder.  The temporal resolution is reduced 32 times, which is done in 5 stages of 2x downsampling block (DownBlock). The temporal information is removed by an average pooling over the temporal dimension. Finally, we use a convolutional layer of kernel size $1$ to project the speaker embedding to the target shape of $d_{\text{spk}}\times1$. In our experiments, we set $d_{\text{spk}}=128$.

Table~\ref{tab:generator} shows the network architecture of the generator. Given an audio of length $T$, the content embedding is of dimension $d_{\text{con}}\times T/256$. The generator uses the content embedding in the main branch. To recover the input temporal resolution using the generator, we upsample the input in 4 stages of 8x, 8x, 2x, and 2x upsampling. A speaker embedding is sampled from the output distribution of the speaker encoder, then it is injected to the generator through all residual blocks.

\begin{table}[!htp]
    \centering
    \caption{Network architecture of the content encoder}
    \label{tab:content_encoder}
    \setlength{\tabcolsep}{0.25em}
    \begin{tabular}{lll}
    \toprule
    Module & Input $\rightarrow$ Output shape & Layer information\\
    \midrule
    Convolution     &  $(1, T) \rightarrow (32, T)$ & Conv(channel=32, kernel=7, stride=1, pad=3)\\
    Residual stack   &  $(32, T)\rightarrow (32, T)$ &  4 x ResidualBlock(channel=32)\\
    Downsample   &  $(32, T)\rightarrow (64, T/2)$ & Conv(channel=64, kernel=4, stride=2, pad=1) \\
    Residual stack   &  $(64, T/2)\rightarrow (64, T/2)$ &   4 x ResidualBlock(channel=64)\\
    Downsample   &  $(64, T/2)\rightarrow (128, T/4)$ & Conv(channel=128, kernel=4, stride=2, pad=1) \\
    Residual stack   &  $(128, T/4)\rightarrow (128, T/4)$ &   4 x ResidualBlock(channel=128)\\
    Downsample   &  $(128, T/4)\rightarrow (256, T/32)$ & Conv(channel=256, kernel=16, stride=8, pad=4) \\
    Residual stack   &  $(256, T/32)\rightarrow (256, T/32)$ &   4 x ResidualBlock(channel=256)\\
    Downsample   &  $(256, T/256)\rightarrow (512, T/256)$ & Conv(channel=512, kernel=16, stride=8, pad=4) \\
    Convolution   &  $(512, T/256)\rightarrow (d_{\text{con}}, T/256)$ & Conv(channel=$d_{\text{con}}$, kernel=7, stride=1, pad=3) \\
    Convolution  &  $(d_{\text{con}}, T/256)\rightarrow (d_{\text{con}}, T/256)$ & Conv(channel=$d_{\text{con}}$, kernel=7, stride=1, pad=3) \\
    \bottomrule
    \end{tabular}
\end{table}

\begin{table}[!htp]
    \centering
    \caption{Network architecture of the speaker encoder}
    \label{tab:speaker_encoder}
    \setlength{\tabcolsep}{0.25em}
    \begin{tabular}{lll}
    \toprule
    Module & Input $\rightarrow$ Output shape & Layer information\\
    \midrule
    Convolution     &  $(80, M) \rightarrow (32, M)$ & Conv(channel=32, kernel=3, stride=1, pad=1)\\
    DownBlock   &  $(32, M)\rightarrow (64, M/2)$ & Conv(channel=64, kernel=3, stride=1, pad=1) \\
    & & + AveragePooling(kernel=2)\\
    DownBlock   &  $(64, M/2)\rightarrow (128, M/4)$ & Conv(channel=64, kernel=3, stride=1, pad=1)  \\
    & & + AveragePooling(kernel=2)\\
    DownBlock   &  $(128, M/4)\rightarrow (256, M/8)$ & Conv(channel=64, kernel=3, stride=1, pad=1) \\
    & & + AveragePooling(kernel=2)\\
    DownBlock   &  $(256, M/8)\rightarrow (512, M/16)$ & Conv(channel=64, kernel=3, stride=1, pad=1) \\
    & & + AveragePooling(kernel=2)\\
    DownBlock   &  $(512, M/16)\rightarrow (512, M/32)$ & Conv(channel=64, kernel=3, stride=1, pad=1) \\
    & & + AveragePooling(kernel=2)\\
    AveragePooling   &  $(512, M/32)\rightarrow (512, 1)$ & AveragePooling(kernel=L/32)  \\
    Mean  &  $(512, 1)\rightarrow (d_{\text{spk}}, 1)$ & Conv(channel=$d_{\text{spk}}$, kernel=1, stride=1, pad=0) \\
    Covariance  &  $(512, 1)\rightarrow (d_{\text{spk}}, 1)$ & Conv(channel=$d_{\text{spk}}$, kernel=1, stride=1, pad=0) \\
    \bottomrule
    \end{tabular}
\end{table}

\begin{table}[!htp]
    \centering
    \caption{Network architecture of the generator}
    \label{tab:generator}
    \setlength{\tabcolsep}{0.25em}
    \begin{tabular}{lll}
    \toprule
    Module & Input $\rightarrow$ Output shape & Layer information\\
    \midrule
    Convolution     &  $(d_{\text{con}}, T/256) \rightarrow (512, T/256)$ & Conv(channel=512, kernel=7, stride=1, pad=3)\\
    Convolution     &  $(512, T/256) \rightarrow (512, T/256)$ & Conv(channel=512, kernel=7, stride=1, pad=3)\\
    Upsample   &  $(512, T/256)\rightarrow (256, T/32)$ & Deconv(channel=256, kernel=16, stride=8, pad=4) \\
    Residual stack   &  $(256, T/32)\rightarrow (256, T/32)$ &  4 x ResidualBlock(channel=256)\\
    Upsample   &  $(256, T/32)\rightarrow (128, T/4)$ & Deconv(channel=128, kernel=16, stride=8, pad=4) \\
    Residual stack   &  $(128, T/4)\rightarrow (128, T/4)$ &  4 x ResidualBlock(channel=256)\\
    Upsample   &  $(128, T/4)\rightarrow (64, T/2)$ & Deconv(channel=64, kernel=4, stride=2, pad=1) \\
    Residual stack   &  $(64, T/2)\rightarrow (64, T/2)$ &  4 x ResidualBlock(channel=64)\\
    Upsample   &  $(64, T/2)\rightarrow (32, T)$ & Deconv(channel=32, kernel=4, stride=2, pad=1) \\
    Residual stack   &  $(32, T)\rightarrow (32, T)$ &  4 x ResidualBlock(channel=32)\\
    Convolution  &  $(32, T)\rightarrow (1, T)$ & Conv(channel=1, kernel=7, stride=1, pad=3) \\
    \bottomrule
    \end{tabular}
\end{table}

\subsection{Details of experimental setups} \label{appexdix:dataset}
\subsubsection{Dataset split and configurations}
All experiments were conducted on the VCTK data set~\citep{vctk_dataset}, which consists of 109 English speakers. From those speakers, six speakers (3 males and 3 females) are randomly extracted as test data for zero-shot voice conversion settings, and the rest used for traditional voice conversion settings. Silences at the beginning and  the end of utterances are trimmed. We use 90\% of the data for training (37,507 utterances) and the rest 10\% for testing (4,234 utterances). Similarly to~\citet{pmlr-v97-qian19c}, a reference of speech ($\sim$20 seconds) is used to compute the speaker embedding for \model{}. 

In traditional voice conversion settings, the objective evaluation is carried out as follows. One conversion is performed per test utterance, in which a target speaker is randomly chosen from the pool of available training speakers. A total of 4,234 conversions have been used to compute the spoofing score. For the subjective evaluation, we randomly choose 3 male and 3 female speakers in the test data. We then perform $6\times 5=30$ conversions from all possible combinations of source-target speaker pairs. There are in total 20 people that participated in our subjective evaluation. 

In zero-shot voice conversion settings, only the subjective test is employed. We use 6 seen speakers and 6 unseen speakers. The experiments are divided into three different settings, including conversions of seen to unseen, unseen to seen, and unseen to unseen. In each of these settings, we report the subjective evaluation of conversions from female to male, female to female, male to female, and female to female.

\subsubsection{Baseline voice conversion methods} \label{appendix:baseline}
For a fair comparison, we use implementations of the baseline methods provided by the corresponding authors\footnote{Links to the implementations of the baseline methods:\\
StarGAN-VC2: \url{https://github.com/SamuelBroughton/StarGAN-Voice-Conversion-2} \\
Blow: \url{https://github.com/joansj/blow} \\
AutoVC: \url{https://github.com/auspicious3000/autovc}} except StarGAN-VC2 (since there is no official implementation from the authors).

\textbf{StarGAN-VC2}~\cite{Kaneko2019} is an improvement of StarGAN-VC~\cite{8639535} with an aim to learn a many-to-many mapping between multiple speakers. StarGAN-VC2 uses the source-and-target conditional adversarial loss, which resolves the issue of unfair training in StarGAN-VC. The improvement in performance is due to the conditional instance normalization~\cite{dumoulin2017learned} that enhances the speaker adaption ability in the generator. However, as noted by~\citet{Karras_2020_CVPR}, instance normalization might cause information loss, which leads to performance degradation.  It is important to note that this method works on the intermediate features, including mel-cepstral coefficients (MCEPs), logarithmic fundamental frequency, and aperiodicities, analyzed by the WORLD vocoder~\cite{Masanori}. The conversion is mainly performed on the MCEPs. The training is based on the least-squares GAN~\cite{8237566}. Compared to our method \model{}, the mapping learned by StarGAN-VC2 is a direct transformation and no disentanglement of the speaker information is employed.

\textbf{Blow}~\cite{NEURIPS2019_9426c311} is an end-to-end method that performs voice conversion directly from the raw audio waveform. The idea is to remove the speaker information when transforming the input space to the latent space. The speaker information is implicitly embedded through the conditional blocks. Inheriting from Glow~\cite{NEURIPS2018_d139db6a}, Blow learns an invertible network architecture with the maximum likelihood objective. The conversion is carried by conditioning directly the weights of the convolutional layers. Blow achieves very competitive results with other state-of-the-art methods. However, the model size is still relatively large compared to that of \model{}. In our experiments, we use the same hyper-parameters as proposed by the authors.

\textbf{AutoVC}~\cite{pmlr-v97-qian19c} achieves the state-of-the-art voice conversion results by disentangling the speaker information from the speech content using a conditional auto-encoder architecture. An information bottleneck is introduced between the encoder and decoder to remove the speaker-identity information. AutoVC has the advantage of implementation simplicity. However, its performance is sensitive to the choice of bottleneck dimension and heavily dependent on the vocoder. In our experiments, we set the bottleneck dimension to 32 and use the WaveNet vocoder~\cite{vandenOord_2016} pre-trained on the VCTK corpus provided by the authors. For zero-shot voice conversion settings, we set the dimension of speaker embedding to 256 and use the speaker encoder model pre-trained on a large corpus provided by the authors. This speaker corpus has in total 3549 speakers, a combination of the VoxCeleb1 and Librispeech data sets.

\subsubsection{Spoofing speaker identification}
Spoofing measure indicates how well a voice conversion method can capture the speaker identities. For this purpose, we train a speaker identification classifier on the same data split used to train the voice conversion method. The classifier follows the network architecture of the speaker encoder (see Appendix~\ref{appendix:network}). However, instead of outputting the parameters of a distribution, the model outputs a linear layer followed by the softmax activation function. Using the cross-entropy loss, we train the classifier with the Adam optimizer using a learning rate of $5\times10^{-4}$ and an exponential scheduler (a decay rate of 0.99) for 150 epochs. On the VCTK corpus, our classifier achieves an accuracy of 99.34\% on the test set of real speech.


\subsection{Additional studies}
\subsubsection{Diversity of outputs}
\begin{figure}[!htp]
    \centering
    \includegraphics[width=0.7\textwidth]{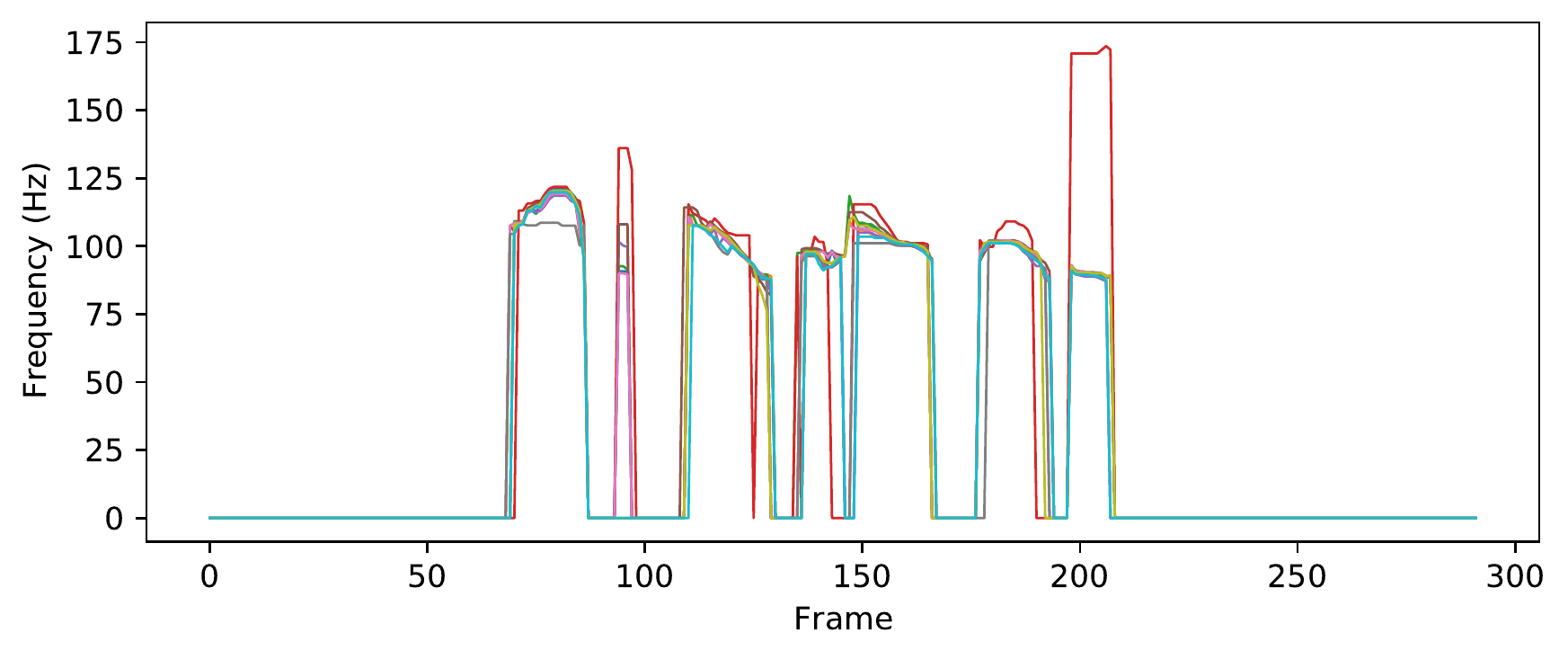}
    \caption{The fundamental frequency (F0) contours of the converted speech}
    \label{appendix:fig:samples}
     \vspace{-2mm}
\end{figure}

\model{} can synthesize diverse samples by changing the latent representation of the speaker embedding. In Fig.~\ref{appendix:fig:samples}, we illustrate the pitch contours of various conversions when sampling different speaker embeddings from the output distribution of the speaker encoder. As shown in the figure, there are various intonation patterns of speech. It is important to note that the VCTK corpus used to train our model has very limited expressive speech.

\subsubsection{Ablation studies} \label{appexdix:ablation}
\textbf{Normalization on content embedding.} To study the effect of using normalization on the content embedding, we conduct an experiment comparing the spoofing results of \model{}$^\dagger$ and \model{} with and without normalization. As shown in Table~\ref{appendix:tab:norm}, employing the normalization on the content embedding helps to improve the performance. We also observe that it reduces the artifacts in the generated audio. Some samples are provided on our demo website \url{https://nvcnet.github.io/}.

\begin{table}[!htp]
 \vspace{-2mm}
   
    \parbox[t]{.45\linewidth}{
     \centering
     \caption{Spoofing results with and without normalization}
    \label{appendix:tab:norm}
    \begin{tabular}{lcc}
    \toprule
      Settings  & \model{}$^{\dagger}$   & \model{} \\
    \midrule
        with  & \textbf{96.43} & \textbf{93.66}  \\
        without  & 95.15  &  91.79 \\
        \bottomrule
    \end{tabular}}
    \hfill
    \parbox[t]{.45\linewidth}{
    \centering
       \caption{Speaker identification results of \model{} with and without random shuffling}
    \label{appendix:tab:rand_shuffle}
    \begin{tabular}{lcc}
    \toprule
      Settings  & Content   & Speaker \\
    \midrule
        with  & \textbf{24.15} & \textbf{99.22}  \\
        without  &  27.91 &  99.17 \\
        \bottomrule
    \end{tabular}
  }
\end{table}
\textbf{Random shuffling.} We study the random shuffle strategy as a data augmentation technique for the speaker encoder. Table~\ref{appendix:tab:rand_shuffle} shows the speaker identification results of \model{} based on the content and speaker embeddings. Large value means speaker information is contained, while small value means less speaker information contained in the representation. For the speaker embedding, the higher value is the better and for the content embedding the lower value is the better. The results show that random shuffling can help to improve the disentanglement of the speaker information from the speech content.

\end{document}